\documentclass[12pt]{article}
\usepackage{a4,amssymb,verbatim,amsmath}
\numberwithin{equation}{section}

\newtheorem{theorem}{Theorem}[section]

\newtheorem{remark}[theorem]{Remark}

\newtheorem{ex}{Example}[section]

\newtheorem{ass}{Assumption}[section]

\numberwithin{equation}{section}

\begin{document}

\newcommand{\ddt}{\partial \over \partial t}
\newcommand{\ddx}{\partial \over \partial x}
\newcommand{\ddu}{\partial \over \partial u}
\newcommand{\ddm}{\partial \over \partial m}

\begin{center}
{\Large {\bf Conservation laws of mean field games equations}}
\end{center}

\begin{center}

{\large Roman Kozlov}

\medskip
\hspace{1.5 cm}

Department of Business and Management Science, Norwegian School of Economics \\
Helleveien 30, 5045, Bergen, Norway; \\
E-mail address: Roman.Kozlov@nhh.no \\

\end{center}


  

\begin{center}
{\bf Abstract}
\end{center}
\begin {quotation}
Mean field games equations are examined for conservation laws. 
The system of mean field games equations consists of two partial differential equations: 
the Hamilton-Jacobi-Bellman  equation for the value function 
and the forward Kolmogorov equation  for the probability density. 
For separable Hamiltonians,  this system has a variational structure, 
i.e.,  the equations of the system are Euler-Lagrange    equations for some Lagrangian functions. 
Therefore, one can use the Noether theorem to derive  the conservation laws 
using  variational and divergence symmetries. 
In order to find such symmetries, we find symmetries of the PDE system 
and select variational and divergence ones.  
The paper considers    separable, state-independent  Hamiltonians 
in one-dimensional state space. 
It examines the most general form of the mean field games system 
for symmetries and conservation laws 
and  identifies particular cases of the system 
which lead to additional symmetries   and conservation laws. 
\end{quotation}

\bigskip















\section{Introduction}

The Partial Differential Equations (PDEs) approach to 
Mean Field Games (MFG) theory is a rapidly developing field of research 
that has its roots in the seminal work of J.-M. Lasry and P.-L. Lions 
in 2006 \cite{Lasry_Lions_2006a, Lasry_Lions_2006b, Lasry_Lions_2007}. 
Another  approach to the development of MFG theory was suggested 
by M. Huang, R. Malhamé, and P. Caines 
\cite{Huang_2006, Huang_2007b}. 
The theory provides a mathematical framework 
for understanding the behaviour of large populations of interacting agents, 
where the entire population influences each agent's behaviour.  
A continuum of agents represents such a population \cite{Aumann_1964}.   
Every again knows the distribution of the other agents 
and uses this to implement its strategy.

The mean field game theory  has found a broad range of applications in various fields  \cite{GLL2011}. 
In economics, MFG theory has been used to study 
market competition 
\cite{Sircar2015, Graber2018, Carmona2021, Chenavaz2021}, 
price formation 
\cite{Gomes2022, Gomes2023}, 
and financial markets 
\cite{Carmona2015, Carmona2017, Fu2021, Fu2023a, Fu2023b}. 
In particular, MFG theory has been applied to understand the behaviour of large populations of investors 
and traders in financial markets 
\cite{Lachapelle2016}. 
MFG models have also been used to study traffic flow and congestion in transportation systems, 
as well as to model the behaviour of crowds in social dynamics 
\cite{Wolfram2011, Wolfram2013, Wolfram2014, Achdou2019}. 
MFG theory has been used to  understand the dynamics of learning 
\cite{Wolfram2016}. 
Additionally, MFG theory has been applied to the study of energy systems, 
such as electricity markets and renewable energy systems 
\cite{Sircar2017, Alasseur2020, Djehiche2020, Tankov2021, Carmona2022}, 
and to the optimization of resource allocation in engineering applications 
\cite{Achdout2016, Achdout2022a}. 
Mathematical aspects 
in different frameworks of the  MFG theory are given 
in several review papers \cite{Gomes2014, Huang_2017}  
and books 
\cite{Carmona2018a, Carmona2018b, Cardaliaguet2019, Cardaliaguet2020}.

Conservation laws play an essential role in differential equations. 
They reflect the  intrinsic properties of the  equations. 
They also help to ensure the consistency and stability of the system. 
In the present paper,  we  analyse the system of mean field games equations for conservation laws. 
This system consists of two PDEs:  
the Hamilton-Jacobi-Bellman (HJB) equation 
for the value function of a typical agent   
and the forward Kolmogorov equation (also known as the Fokker-Plank equation)  
for the evolution of the density of the agent population.    
It should be noted that the first equation is backward in time, and the latter is forward in time.   
There are several approaches to derive conservation laws \cite{Naz2008}. 
In the paper, we exploit the variational structure of the MFG equations. 
For separable Hamiltonians,  this system has a variational structure, 
i.e.,  the equations of the system are Euler-Lagrange    equations for some Lagrangian functions. 
Therefore, one can use symmetries   
and the Noether theorem \cite{bk:Noether1918} to derive the conservation laws.

Lie point symmetries provide efficient tools 
to analyse {non}linear differential equations
\cite{bk:Ovsiannikov1978, bk:Ibragimov1985, bk:Olver}. 
The symmetries  are related to essential properties of the models 
represented by the differential equations:

\begin{itemize}

\item 

Symmetries generate transformations 
which transfer solutions of the  differential equations  into other solutions.  
In particular, they can be used to find new solutions from the known ones.

\item 

Symmetries of PDEs can  be used to find particular solutions. 
These solutions are called  invariant solutions.

\item 

Symmetries of variational differential equations can be used for Noether’s theorem, 
which provides conservation laws. 

\end{itemize}

For the application of the Noether theorem, 
one needs variational formulation  of the considered differential equations 
and variational or divergence symmetries. 
Such symmetries are also symmetries of the  considered  differential equations. 
Therefore, the standard approach is to find  symmetries of the   differential equations, 
select those that are also variational or divergence symmetries, 
and apply the Noether theorem to obtain  conservation laws. 
These steps are implemented in the paper. 
In particular, we start with the most general case of the differential equations, 
for which there are a few symmetries, 
and continue with identifying particular cases, 
for which there are additional symmetries. 
After that,  we select symmetries suitable for the Noether theorem 
and  provide the  conservation law.

The paper is organized as follows. 
In the next Section we review the mean field games equations 
and specify the cases with the variational formulation. 
Section \ref{Symmetries} describes Lie point symmetries and the Noether theorem. 
This section is written for the particular system   which is analysed 
in Sections  \ref{main_section}  and  \ref{Conservation_laws}, 
where we present the symmetry analysis and the conservation laws. 
Section \ref{Interpretation} discusses the interpretation of the obtained conservation laws.  
Final Section \ref{Conclusion} provides concluding remarks. 
Some technical details are extracted into the Appendices.

\section{Variational MFG equations}

\label{Background}



The system of mean field games equations consists of two coupled PDEs: 
the Hamilton-Jacobi-Bellman  equation  
and the forward Kolmogorov equation  \cite{Cardaliaguet2020}: 
\begin{subequations}   \label{general_system}
\begin{gather}     \label{general_equation_1} 
- u_t  -  \varepsilon  \Delta   u + H ( x, \nabla u , m ) = 0  
\qquad
\mbox{in}  \   [ 0, T] \times  \mathbf{R}^d , 
\\
  \label{general_equation_2} 
   m_t  -  \varepsilon  \Delta   m   -  \mbox{div} ( m H _{\nabla u}  ( x, \nabla u , m )  ) = 0  
\qquad
\mbox{in}  \    [ 0, T]   \times  \mathbf{R}^d . 
\end{gather}
\end{subequations}
The scalar functions $ u ( t,x) $ and  $ m( t,x) $ 
stand for the value function 
and the probability density. 
They depend on time $ t \in [ 0,T]$ 
and spatial independent variable $  x \in  \mathbf{R}^d $.  
$ H  ( x, \nabla u , m )  $ is the Hamiltonian associated to the game 
and $ H _{\nabla u}  =  (  H _{u_{x_1}}  , \cdots , H _{u_{x_d}}   ) $.   
Parameter   $ \varepsilon \geq 0 $ characterizes the noise.     
It should be noted that the HJB equation    (\ref{general_equation_1}) 
is backward in time and 
the Kolmogorov equation   (\ref{general_equation_2})  
is forward in time. 
Thus,  the initial-terminal conditions are given as  
\begin{equation}
m(0 , x) =  m_0 (x) , 
\qquad 
u (T , x)  = G ( x , m(T,x) ) 
\qquad
\mbox{in}  \    \mathbf{R}^d . 
\end{equation}
Since $ m( t,x) \geq 0 $  is the probability density, we have 
\begin{equation}    \label{normalization}
\int m dx =1 .
\end{equation}

The solution $ (u,m) $  of the coupled system  (\ref{general_system})  
represents  a Nash equilibria of the mean field game. 
We refer to    \cite{Cardaliaguet2020} 
for detailed assumptions 
one needs for the existence and uniqueness of the solutions. 
A sketch of the derivation of the system  
is given in Appendix   \ref{HJB-KE}.


\subsection{Separable Hamiltonians}

A particular case of the system   (\ref{general_system})  
\begin{subequations}   \label{separable_system}
\begin{gather}
- u_t  -  \varepsilon  \Delta   u + H ( x, \nabla u  )  -    f ( x , m )  = 0     , 
\\
   m_t  -  \varepsilon  \Delta   m   -  \mbox{div} ( m H _{\nabla u}  ( x,  \nabla u  )  )  = 0   
\end{gather}
\end{subequations}
corresponds to separate  Hamiltonians 
 $   H ( x, \nabla u  , m )  =  H ( x, \nabla u  )  -    f ( x , m ) $.  
This system has a variational formulation. 
It  is provided by the Lagrangian function
\begin{equation}      \label{separable_Lagrangian}
L =    - m u_t  
+   \varepsilon     \nabla m     \cdot     \nabla u  
+ m  H ( x, \nabla u  ) 
- F  ( x , m )  , 
\end{equation}
where 
\begin{equation}
F  ( x , m )   =\int   f ( x , m )  d m , 
\end{equation}
i.e.,  equations   (\ref{separable_system}) 
are Euler-Lagrange  equations  
\begin{subequations}   \label{Lagrangian_form}
\begin{gather}
{ \delta L \over \delta m }  
=  { \partial  L \over \partial m }  
- D_t   { \partial  L \over \partial m_t  }  
-  \sum _ {i=1} ^d    
D_{x_i}     { \partial  L \over \partial   m _{x_i}    }  
= 0  , 
\\
{ \delta L \over \delta u }  
=  { \partial  L \over \partial u }  
- D_t   { \partial  L \over \partial u_t  }  
-  \sum _ {i=1} ^d    
 D_{x_i}    { \partial  L \over \partial   u_{x_i}  }  
= 0
\end{gather}
\end{subequations}
for Lagrangian     (\ref{separable_Lagrangian}). 
Here  $ D_t $ and $ D_{x_i}  $ are total differentiation operators 
for time $t$  and spatial variables $ {x_i}  $, respectively. 

\subsection{One-dimensional state space  with state-independent Hamiltonians.}  

In this paper, we examine the simplest case of system  (\ref{separable_system}): 
the case of one-dimensional state space  $ d =1 $ 
with  state-independent  functions $H = H  ( u_x) $  and $ f = f(m) $. 
The equations take the form  
\begin{subequations}     \label{system_mu}
\begin{gather}    \label{equation_u}
F_1   = - u_t  -  \varepsilon   u _{xx}  + H ( u_x )  -  f ( m ) = 0   
\qquad
\mbox{in}  \  [ 0, T]   \times  \mathbf{R} , 
\\
  \label{equation_m}
F_2   =    m_t  -  \varepsilon  m _{xx}   -   m _x H '  ( u_x )     -      m H ''  ( u_x )  u _{xx}     = 0 
\qquad
\mbox{in}  \   [ 0, T]  \times  \mathbf{R} . 
\end{gather}
\end{subequations}
We consider second-order MFG systems  restricting to   $ \varepsilon > 0  $. 
Assumptions 
\begin{equation}     \label{Assumptions}
H '' ( u_x ) {\not \equiv}  0 , 
\qquad 
f ' (m)   {\not \equiv}  0  ; 
\end{equation} 
guarantee that the equations are genuinely coupled, 
i.e., none of these equations can be considered independently from the other.

Equations    (\ref{system_mu}) 
are Euler-Lagrange equations for the Lagrangian  function 
\begin{equation}     \label{Lagrangian}
L =    - m u_t  
+   \varepsilon     m _x   u _x    
+ m  H ( u_x ) 
- F  ( m )  
\end{equation}
with 
\begin{equation}    \label{Function_F}
F  (  m )   =\int   f (  m )  d m . 
\end{equation}

In Section  \ref{main_section}, 
we analyse the system of equations   (\ref{system_mu}) 
for admitted symmetries 
and in Section \ref{Conservation_laws} 
we derive the conservation laws of these equations.

\subsection{Other variational cases}

Between the most general variational system   (\ref{separable_system}) 
and the simplest variational  system  (\ref{system_mu}),  
two particular MFG systems might be worthy of a separate study.   
They correspond to

\begin{itemize}

\item 

Multi-dimensional state space   $  x \in \mathbf{R}^d $   with  state-independent  Hamiltonians. 

For state-independent functions  $H =  H ( \nabla u  )  $ and $f = f(m) $, 
we get the particular case of the system (\ref{separable_system}) 
\begin{subequations}   \label{x_independent}
\begin{gather}
   - u_t  -  \varepsilon  \Delta   u + H ( \nabla u  ) -  f ( m )  = 0 
\qquad
\mbox{in}  \    [ 0, T]   \times  \mathbf{R}^d , 
\\
    m_t  -  \varepsilon  \Delta   m   -  \mbox{div} ( m H _{ \nabla u }    ( \nabla u  )  )  = 0  
\qquad
\mbox{in}  \    [ 0, T]    \times  \mathbf{R}^d , 
\end{gather}
\end{subequations}
provided by Lagrangian 
\begin{equation*}     
L =   - m u_t 
+   \varepsilon    \nabla m      \cdot      \nabla u   
+ m  H ( \nabla u  ) 
- F  ( m )  
\end{equation*}
with $ F (m) $ given by (\ref{Function_F}).

\item

One-dimensional state space

If the  state space is one-dimensional, 
system  (\ref{separable_system}) gets simplified as 
\begin{subequations}   \label{one_dim_system}
\begin{gather}
- u_t  -  \varepsilon   u _{xx}  + H ( x, u_x ) -  f ( x , m ) = 0   
\qquad
\mbox{in}  \   [ 0, T]   \times  \mathbf{R} , 
\\
   m_t  -  \varepsilon  m _{xx}   -    ( m H _{u_x}  ( x, u_x )  ) _x  = 0   
\qquad
\mbox{in}  \   [ 0, T]   \times  \mathbf{R}  . 
\end{gather}
\end{subequations}
It's Lagrangian function is 
\begin{equation*}    
L =    - m u_t 
+   \varepsilon        m_x   u_x     
+ m  H ( x,  u _ x ) 
- F  ( x , m )  . 
\end{equation*}

\end{itemize}


\section{Lie point symmetries and the Noether theorem}

\label{Symmetries}

This section reviews  Lie point symmetries and the Noether theorem 
\cite{bk:Ovsiannikov1978, bk:Ibragimov1985, bk:Olver}. 
For convenience, 
the discussion is adopted to the system of PDEs   (\ref{system_mu}) 
and first-order Lagrangian function    (\ref{Lagrangian}), 
which are  analysed in this paper.

Lie point symmetries of system (\ref{system_mu}) are given by generators of the form 
\begin{equation}    \label{generator}
X 
=  \xi ^t  {\ddt} 
+  \xi ^x  {\ddx} 
+  \eta  ^u  {\ddu}  
+  \eta  ^m  {\ddm}  
\end{equation}
with generator coefficients 
\begin{equation*}
  \xi ^t  =    \xi ^t  ( t,x,m,u) , 
\qquad 
  \xi ^x =    \xi ^x  ( t,x,m,u) , 
\qquad
 \eta  ^u  =   \eta  ^u   ( t,x,m,u) , 
\qquad 
 \eta  ^m  =  \eta  ^m   ( t,x,m,u) .  
\end{equation*}
For application to differential equations  
a generator should  be prolonged to all variables 
involved  in the considered differential equations. 
In our case,  we get the prolonged generator
\begin{multline}       \label{prolonged_generator}
X 
=  \xi ^t  {\ddt} 
+  \xi ^x  {\ddx} 
+  \eta  ^u  {\ddu}  
+  \eta  ^m  {\ddm}  
\\
+  \zeta _t  ^u  {\partial \over \partial u_t}   
+  \zeta _t  ^m  {\partial \over \partial m_t}   
+  \zeta _x  ^u  {\partial \over \partial u_x}   
+  \zeta _x  ^m  {\partial \over \partial m_x}   
+  \zeta _{xx}  ^u  {\partial \over \partial u _{xx} }   
+  \zeta _{xx} ^m  {\partial \over \partial m _{xx} }   . 
\end{multline}
The coefficients for the derivatives are computed  
according to the standard prolongation  formulas 
\cite{bk:Ovsiannikov1978, bk:Ibragimov1985, bk:Olver}
\begin{equation*}
\zeta _t  ^u  = D_t ( \eta  ^u  )   - u_t  D_t (  \xi ^t )   - u_x  D_t (  \xi ^x ) , 
\qquad 
\zeta _t  ^m  = D_t ( \eta  ^m  )   - m_t  D_t (  \xi ^t )   - m_x  D_t (  \xi ^x )  , 
\end{equation*}
\begin{equation*}
\zeta _x  ^u  = D_x ( \eta  ^u  )   - u_t  D_x (  \xi ^t )   - u_x  D_x (  \xi ^x )   , 
\qquad 
\zeta _x  ^m  = D_x ( \eta  ^m  )   - m_t  D_x (  \xi ^t )   - m_x  D_x (  \xi ^x )  , 
\end{equation*}
\begin{equation*}
\zeta _{xx}  ^u  = D_x (  \zeta _x  ^u   )   - u _{tx}  D_x (  \xi ^t )   - u _{xx}  D_x (  \xi ^x )  , 
\qquad 
\zeta _{xx}  ^m  = D_x ( \zeta _x  ^u   )   - m _{tx}  D_x (  \xi ^t )   - m _{xx}  D_x (  \xi ^x ) , 
\end{equation*}
where  $ D_t $  and $ D_x $ are total differentiation operators for independent  variables $t$ and $x$, 
respectively, 
i.e. 
\begin{equation*}
D_t =  {\ddt}  
+  u_t   {\ddu}  
+  m_t   {\ddm}  
+ \cdots , 
\qquad 
D_x =  {\ddx}  
+  u_x   {\ddu}  
+  m_x   {\ddm}  
+ \cdots 
\end{equation*}

To find Lie point symmetries admitted by differential equations, 
one applies the   infinitesimal invariance criterion  \cite{bk:Ovsiannikov1978, bk:Ibragimov1985, bk:Olver}.  
It  requires that 
 the action of the prolonged generator 
on the  considered differential  equations 
provides identities on the solutions of these equations.  
For system (\ref{system_mu}),   the   infinitesimal invariance criterion takes the form 
\begin{subequations}        \label{infinitesimal}
\begin{gather}
E_1  = 
\left. 
X ( F_1  )
\right| _{  F_1 = 0 , F_2 =0 } 
= 0 , 
\\
E_2 = 
 \left. 
 X (  F_2   )
 \right| _{  F_1 = 0 , F_2 =0 } 
= 0  .
\end{gather}
\end{subequations}  
These equations are PDEs for the coefficients of the generator   (\ref{generator}). 
Solving them, one finds the admitted generators.

Lie point symmetries of the differential  equations are needed for the Noether theorem.

\begin{theorem}  (E. Noether  \cite{bk:Noether1918}) 
Let Lagrangian function   (\ref{Lagrangian})   satisfy  the equation
\begin{equation}    \label{variational}
X L + L ( D_t (  \xi ^t ) +  D_x (  \xi ^x )  ) =  0 
\end{equation}
for generator $ X$ given by  (\ref{generator}), 
then the generator $X$ is a symmetry 
of the system of Euler-Lagrange equations (\ref{system_mu})
and there holds the conservation law
\begin{equation}
D_t   [  T^t  ]   +    D_x   [  T^x  ]   = 0 , 
\end{equation}
where 
\begin{subequations}     \label{components} 
\begin{gather} 
 \label{component1} 
T^t  
= \xi^t  L 
+ (  \eta  ^u     -    \xi ^t u_t    -   \xi ^x   u_x   )    {\partial L \over \partial u_t} 
+ (  \eta  ^m     -    \xi ^t m_t    -   \xi ^x   m_x   )    {\partial L \over \partial m_t}   , 
\\
   \label{component2} 
T^x  
= \xi^x  L 
+ (  \eta  ^u     -    \xi ^t u_t    -   \xi ^x   u_x   )    {\partial L \over \partial u_x} 
+ (  \eta  ^m     -    \xi ^t m_t    -   \xi ^x   m_x   )    {\partial L \over \partial m_x} . 
\end{gather}
\end{subequations}
\end{theorem}

There is a natural extension of the Noether theorem.

\begin{remark}  (\cite{Bessel-Hagen}) 
If instead of relation   (\ref{variational}), there holds 
\begin{equation}    \label{divergence}
X L + L ( D_t (  \xi ^t ) +  D_x (  \xi ^x )  ) =  D_t (  V ^t ) +  D_x (  V ^x ) 
\end{equation}
with some functions  $ V ^t ( t, x, u, m)  $   and   $ V ^x ( t, x, u, m)  $, 
then the generator $X$ is a symmetry 
of the system of Euler-Lagrange equations (\ref{system_mu})
and there holds the conservation law
\begin{equation}
D_t   [  T^t  - V ^t  ]   +    D_x   [  T^x - V ^x ]   = 0 
\end{equation}
with $T^t$ and $T^x$ given by   (\ref{components}).  
\end{remark}

Symmetries of the variational equations 
which satisfy (\ref{variational}) are called {\it variational} symmetries 
and 
symmetries with property  (\ref{divergence})  are called {\it divergence} symmetries.

\section{Symmetry analysis of system    (\ref{system_mu})}

\label{main_section}

This section provides  analysis of system   (\ref{system_mu}),(\ref{Assumptions}) 
for admitted  symmetries.  
Lie point symmetries of system (\ref{system_mu})  are provided 
by the infinitesimal invariance criterion  (\ref{infinitesimal}). 
In detail, these equations take the form 
\begin{equation*}      
E_1 = 
\left[
 -  \zeta _t  ^u  
-  \varepsilon    \zeta _{xx}  ^u  + 
H' (  u_x )  \zeta _x  ^u 
 -  f '  ( m )   \eta  ^m  
\right]
_{    F_1 = 0 ,  F_2 = 0     } 
= 0 , 
\end{equation*} 
\begin{multline*} 
E_2 = 
\left[
\zeta _t  ^m
 -  \varepsilon  \zeta _{xx} ^m  
-  \zeta _x  ^m    H '  ( u_x )  
-   m _x   H ''  ( u_x ) \zeta _x  ^u    
 \right.
 \\ 
\left. 
-    \eta  ^m  H ''  ( u_x )  u _{xx}
- m  H '''  ( u_x )  \zeta _x  ^u    u _{xx}
-   m H ''  ( u_x )  \zeta _{xx}  ^u 
\right]
_{   F_1 = 0 ,  F_2 = 0     } 
= 0  .
\end{multline*}  
These equations are used  to find symmetries 
admitted for arbitrary $H (u_x) $  and $ f (m) $. 
They are also used to identify special cases of   $H (u_x) $  and $ f (m) $ 
which lead to extensions of the symmetries admitted for arbitrary $H (u_x) $  and $ f (m) $.  
The preliminary analysis of these equations aimed at 
identifying the special cases of the Hamiltonian function is given in  Appendix \ref{preliminary}. 
It results in four special cases of $H (u_x) $ 
which should be analysed in addition to the general case of arbitrary  $H (u_x) $.  
Thus, we have the following cases to examine

\begin{itemize}

\item 

General case: 
arbitrary 
$  H  (u_x )  $

\item 

Special cases:

\begin{itemize}

\item 

power 
\begin{equation}     \label{form_2}
 H  (u_x )  =      h  (  u_x + q ) ^p     +  h_2  u_x ^2  +  h_1  u_x   +  h_0  , 
\qquad 
h \neq 0 ,  
\qquad 
p  \neq 0, 1, 2, 3  ; 
\end{equation}

\item 
exponential 
\begin{equation}     \label{form_3}
 H  (u_x )  =      h   e^{ k  u_x }   +  h_2  u_x ^2  +  h_1  u_x   +  h_0  , 
\qquad 
h \neq 0 ,  
\qquad 
k  \neq 0  ; 
\end{equation}

\item cubic (particular power) 
\begin{equation}     \label{form_4}
 H  (u_x )  =     h  u_x ^3      +  h_2  u_x ^2  +  h_1  u_x   +  h_0  , 
\qquad 
h \neq 0  ; 
\end{equation}

\item quadratic (particular power) 

\begin{equation}     \label{form_5}
 H  (u_x )  =      h  u_x ^2  +  h_1  u_x   +  h_0  , 
\qquad 
h \neq 0  .
\end{equation}

\end{itemize}
Here $ h$, $ h_2$, $ h_1$, $ h_0$, $ p$, $ q$  and $ k$   are constants.   

\end{itemize}

Before we consider these cases,
it is appropriate to simplify them 
by transformations  which do not change the form of equations (\ref{system_mu}). 
Such transformations are called {\it equivalence transformations}.  
We apply the following steps.

\begin{enumerate}

\item

Transformations of the form 
\begin{equation}    \label{transformation_1}
u  \rightarrow   \bar{u}  
= u   + A x   , 
\qquad 
A = \mbox{const}
\end{equation}
are used to remove  parameter $q$  in Hamiltonian  (\ref{form_2}),  
to scale coefficient $h$ in  Hamiltonian  (\ref{form_3}) 
and to remove the term $h_2 u_x ^2 $ in Hamiltonian  (\ref{form_4}).

\item

Transformations 
\begin{equation}    \label{transformation_2}
u  \rightarrow   \bar{u}  
=   u + B t, 
\qquad 
B = \mbox{const}
\end{equation}
are used to remove  constant terms $ h_0 $ in all Hamiltonians     (\ref{form_2})-(\ref{form_5})
as well as in functions $f(m)$  (if $f(m)$ contains a  constant term).

\item 

The transformation
\begin{equation}      \label{transformation_3}
x  \rightarrow   \bar{x}  
=  x  +  h_1 t   
\end{equation}
removes the terms $h_1u_x  $  in all Hamiltonians   (\ref{form_2})-(\ref{form_5}).

\item

Scaling of independent and dependent variables 
\begin{equation}      \label{transformation_4}
t  \rightarrow   \bar{t} 
=   C_1  t   , 
\qquad 
x  \rightarrow   \bar{x}
=   C_2  x  , 
\qquad 
u  \rightarrow   \bar{u}
=    C_3 u  ,    
\qquad
m  \rightarrow   \bar{m}
=   C_4 m , 
\end{equation}
where   $ C_1$, $C_2 $, $C_3$ and $C_4$  are constants such that $     C_1  C_2 C_3 C_4 \neq 0 $,  
are used to make a suitable choice of the leading term coefficient $h$  in Hamiltonians  
(\ref{form_2}),   (\ref{form_4}) and  (\ref{form_5}).  
Scaling  transformations  can be chosen so that they satisfy 
\begin{equation*}  
C_2 C_4 =1 . 
\end{equation*}
In addition to the form of equations (\ref{system_mu}) 
such  scaling  transformations  preserve normalization (\ref{normalization}). 
Function $f(m)$ is not specified,  and it can be changed 
by the scaling transformations.   
\end{enumerate}

Implementing these steps, we arrive at the following  simplified 
Hamiltonians for the special cases:

\begin{itemize} 

\item 

power 
\begin{equation}     \label{form_2_mod}
 H  (u_x )  =      {  1\over p}   u_x ^p     +  h_2  u_x ^2  , 
\qquad 
p  \neq 1, 2, 3 ; 
\end{equation}

\item 

exponential 
\begin{equation}     \label{form_3_mod}
 H  (u_x )  =      {  1  \over k}    e^{ k  u_x }   +  h_2  u_x ^2    ; 
\end{equation}

\item 

cubic 
\begin{equation}     \label{form_4_mod}
 H  (u_x )  =  {  1  \over 3 }   u_x ^3       ; 
\end{equation}

\item 

quadratic 
\begin{equation}     \label{form_5_mod}
 H  (u_x )  =   {  1  \over 2 } u_x ^2   .
\end{equation}

\end{itemize} 
Here $h_2$  is an arbitrary constant.

Detailed analysis of determining equations    (\ref{infinitesimal}) 
for the general  and special cases of the Hamiltonian 
provides the following result.

\begin{theorem}     \label{theorem-symmetries} 
Up to equivalence transformations (\ref{transformation_1})--(\ref{transformation_4}),  
system  (\ref{system_mu}),(\ref{Assumptions}). 
has the following cases of admitted symmetries:

For arbitrary $ H (u_x) $ and arbitrary $ f(m) $,  the system admits the three symmetries 
\begin{equation}    \label{basis}
X_1 = {\ddt}  , 
\qquad
X_2 = {\ddx} , 
\qquad 
X_3 = {\ddu}      . 
\end{equation}
If  $ f(m)   = \alpha \ln m  $,  
there exists the additional symmetry 
\begin{equation}    \label{basis_plus}
X_f =  \alpha  t  {\ddu}    - m   {\ddm}  . 
\end{equation}

There are particular cases of $ H(u_x)$ 
which lead to the extensions of the admitted symmetry  group.



\noindent {\bf a)} 
For power $ H (u_x) =  {  1\over p}   u_x ^p  $ 
(Hamiltonian (\ref{form_2_mod}) with  $ h_2 = 0$)  
and   power $ f(m)  =   \alpha  m^{\gamma} $,   
there is the additional generator 
\begin{equation}        \label{special_2}
X  ^a  _4
=   2 (p-1) t {\ddt}
+ (p-1)  x {\ddx} 
+ (p-2)  u   {\ddu}
-  { p \over \gamma  } m   {\ddm} .
\end{equation}


\noindent {\bf b)} 
For exponential    $ H (u_x)  =  {  1  \over k}    e^{ k  u_x } $ 
(Hamiltonian (\ref{form_3_mod}) with  $ h_2 = 0$)  
and   power $ f(m)  =   \alpha  m^{\gamma}  $, 
there exists  the additional symmetry 
\begin{equation}     \label{special_3}
X  ^b _4 
=   2 t {\ddt}
+   x {\ddx} 
+ \left(  u  - { x \over k} \right)    {\ddu}
-  { 1 \over \gamma  } m   {\ddm} . 
\end{equation}


\noindent {\bf c)} 
For quadratic   $ H (u_x)  = {  1\over 2}   u_x ^2  $  and arbitrary  $ f(m)  $ 
there is the generator 
\begin{equation}    \label{special_5a}
X  ^c 
=   t {\ddx} 
-  x      {\ddu}  . 
\end{equation}
If    $ f(m)  =   \alpha  m^{2} $,  then there is the particular case of symmetry    (\ref{special_2}), 
corresponding the  $p=2$, namely
\begin{equation}        \label{special_5_2}
X  ^c _4 
= 
\left. 
X  ^a  _4
\right| _ {p=2,   \gamma =2}
=   2  t {\ddt}
+  x {\ddx} 
-   m   {\ddm}  , 
\end{equation}
and one more symmetry  
\begin{equation}     \label{special_5b}
X  ^c  _5 
=   t ^2  {\ddt}
+   t x {\ddx} 
+ \left(   - { x^2  \over 2 }   + \varepsilon  t   \right)    {\ddu}
-    t   m   {\ddm}  .
\end{equation}


\end{theorem}

\noindent {\it Proof.}
The system of  determining equations    (\ref{infinitesimal})  is an overdetermined system of PDEs 
for the coefficients of generator $X$,  given by (\ref{generator}).   
These equations are analysed for the case of arbitrary Hamiltonian   $ H (u_x)  $ 
and for four special  Hamiltonians   (\ref{form_2_mod})-(\ref{form_5_mod}). 
For the considered cases of the Hamiltonian function 
the determining equations   split (first for the derivatives of $u$ and $m$, then for the variables) 
into equations that specify the symmetry coefficients.

Particular cases of $f(m)$, namely 
$ f(m)   = \alpha \ln m  $,  $ f(m)  =   \alpha  m^{\gamma} $ and $ f(m)  =   \alpha  m^{2} $,  
are identified from the detailed analysis of the    determining equations 
as cases that provide additional symmetries. 
 \hfill $\Box$

\medskip

The admitted symmetries can be  divided into the basis symmetry groups, 
which is admitted for all functions   $ H (u_x)  $  and  $f(m)$, 
and the additional symmetries, 
which exist for the particular  functions   $ H (u_x)  $  and  $f(m)$. 
The basis symmetry group consists of three generators  $\{ X_1, X_2, X_3\} $, 
 given in   (\ref{basis}). 
For convenience, we present the extensions of the basic symmetry group in Table~1.

\begin{equation*}
\begin{array}{|c|c|c|c|c|}
\hline
   &   
\mbox{arbitrary $f(m)$}    &    
f(m)   = \alpha \ln m   &  
f(m)  =   \alpha  m^{\gamma} , 
\ 
\gamma \neq 2 
& f(m)  =   \alpha  m^{2} 
\\
\hline
\mbox{arbitrary $H  (u_x )$}   &    
 -   &   
X_f  & 
-   &        
 - \\
\hline
H  (u_x )  =      {  1\over p}   u_x ^p     & 
 -   &   
X_f  & 
X^a _4   &    
\left. X^a _4 \right| _{\gamma=2} 
\\
\hline
H  (u_x )  =      {  1  \over k}    e^{ k  u_x }  &  
 -  &   
X_f  & 
X_4 ^b    &    
\left. X^b _4 \right| _{\gamma=2} 
\\
\hline
 H  (u_x )  =   {  1  \over 2 } u_x ^2     &  
  X  ^c   &   
X_f , X ^c   & 
X ^c   , X_4   &    
X ^c  , X ^c _4 ,  X ^c _5 
\\
\hline
\end{array}
\end{equation*}
\begin{center}
{\bf Table 1.} Extensions of the basic symmetry group  with generators  $\{ X_1, X_2, X_3\} $, 
which is  admitted for arbitrary $H(u_x)$ and  arbitrary $ f(m)$.  
Classification is performed up to equivalence transformations (\ref{transformation_1})--(\ref{transformation_4}). 
\end{center}

\begin{remark}
In Theorem   \ref{theorem-symmetries}, 
the case of cubic Hamiltonian   (\ref{form_4_mod}) 
stands as a particular case of power Hamiltonian (\ref{form_2_mod}). 
The cubic Hamiltonian was identified separately from the  power Hamiltonian 
in the preliminary analysis of the determining equations, given in Appendix \ref{preliminary}.  
However, it has the same symmetry properties as the general  power Hamiltonian  case.  
\end{remark}

\begin{remark}
It is possible to employ scaling transformation  to get rid of the parameter   $ \alpha $ 
in the particular cases of $ f(m) $  
without  changing the form of equations  (\ref{system_mu}). 
However, for some parameter values, 
such scaling transformations   do not keep normalization relation (\ref{normalization}). 
For this reason, parameter  $ \alpha $ is kept. 
\end{remark}

\section{Conservation laws of system   (\ref{system_mu})}

\label{Conservation_laws}

The cases of $H$  and $f$, which should be considered, are specified by   Theorem  \ref{theorem-symmetries}. 
It remains to select variational and divergence symmetries 
and apply the Noether theorem to derive the conservation laws.

\subsection{General case: arbitrary  $ H (u_x) $}

For arbitrary Hamiltonian, two cases of the function $ f(m) $ exist to consider.

\medskip

\noindent   {\bf 1)}    Arbitrary $ f(m) $ 

In this case, we consider 
\begin{equation}          \label{Arbitrary_f}
F(m) = \int f(m) dm 
\end{equation}
corresponding  to arbitrary $ f(m) $.  

Equations   (\ref{system_mu}) admit symmetries $ X_1 $,  $ X_2 $  and $ X_3 $,  
given in (\ref{basis}).  
These three symmetries are variational.

\begin{itemize}

\item

The symmetry   $ X_1 $ provides the conservation law
\begin{equation}      \label{CLG1} 
D_t \left[ 
   \varepsilon     m _x    u _x    
+ m  H ( u_x ) 
- F  ( m )  
\right]
- 
D_x \left[ 
   \varepsilon   (   u _t      m _x  +   m_t   u _x   ) 
+  m u_t  H ' ( u_x ) 
\right] 
= 0   . 
\end{equation}

\item 
 
The symmetry $ X_2 $ leads to  the conservation law
\begin{equation}      \label{CLG2} 
D_t \left[ 
 m u_x  
\right]
- 
D_x \left[ 
  m u_t 
+    \varepsilon   m _x   u _x    
+ m (  u_x H '  (u_x)   - H ( u_x )  )
+  F  ( m )  
\right]  
= 0  . 
\end{equation}

\item 

The symmetry  $ X_3 $ 
gives the conservation law
\begin{equation}       \label{CLG3} 
- D_t \left[ 
   m  
\right]  
+ 
D_x \left[ 
\varepsilon   m _x   
+  m  H ' ( u_x ) 
\right]  
= 0 .  
\end{equation}
Note that this conservation law is the Kolmogorov equation    (\ref{equation_m}) itself. 
It reflects the conservation of the  probability density  $ m(t,x) $.

\end{itemize}


\medskip

\noindent   {\bf 2)}  
  Logarithmic    $ f(m)   = \alpha \ln m $ 

In this case 
\begin{equation}
F(m) =  \alpha   ( m \ln m  - m )   . 
\end{equation}
The generator $ X_f $ (see  (\ref{basis_plus})) is 
neither variational nor divergence symmetry. 
Thus, it does not provide a conservation law.

\subsection{Special cases of  $H (u_x) $}

There are three exceptional cases of Hamiltonian: 
(general) power, exponential and quadratic.



\medskip

\noindent {\bf a)} 
Power $ H (u_x) =  {  1\over p}   u_x ^p $ 
and power  $ f(m)  =   \alpha  m^{\gamma} $  

In this case 
\begin{equation}    \label{F_case_2}
F(m)  =  {   \alpha   \over  \gamma+1 }   m^{\gamma+1}  . 
\end{equation}
Symmetry   $ X  ^a _4 $, given by (\ref{special_2}), 
 is variational,  provided that 
\begin{equation}
\gamma = \gamma_* = { p \over  2 p -3  }   . 
\end{equation}
Thus we consider the  symmetry 
\begin{equation}      \label{special_2_specified}
{Y} ^a _4 
=
\left.
X_4  
\right| _{ \gamma = {\gamma _*  } } 
=   2 (p-1) t {\ddt}
+ (p-1)  x {\ddx} 
+ (p-2)  u   {\ddu}
-  ( 2 p - 3 )  m   {\ddm}   . 
\end{equation}
It provides the conservation laws
\begin{multline}      \label{CL4a} 
D_t \left[ 
 2 ( p -1 )  t 
\left( 
\varepsilon       m _x  u _x    
+    {  1\over p}  m   u_x ^p   
-   { \alpha   \over  \gamma _*  +1 }   m^{\gamma  _*  +1}     
\right)
+   ( p -1 )
 x  m u_x  
-  
 ( p - 2 )  
m u 
\right]  
\\ 
-  
D_x \left[ 
  2 ( p-1 )  t 
\left( 
 \varepsilon   (   u _t      m _x   +    m_t   u _x  ) 
+  m u_t   u_x ^{p-1} 
  \right)
\phantom{7 \over 7}
\right. 
\\
+    ( p-1  ) x 
\left( 
 m u_t  
+    \varepsilon   m _x   u _x    
+   { p - 1  \over p }  m u_x ^p 
+  { \alpha   \over  \gamma _* +1 }   m^{\gamma _* +1}  
\right)
\\
\left.  
\phantom{51 \over 51}
-   ( p - 2 )  u (      \varepsilon     m _x   +    m   u_x ^{p-1}  ) 
+ ( 2 p -3  )   \varepsilon   m   u _x     
\right]  
= 0  . 
\end{multline}


\medskip

\noindent {\bf b)} 
 Exponential    $ H (u_x) =  {  1  \over k}    e^{ k  u_x }   $ 
and  power  $ f(m)  =   \alpha  m^{\gamma} $  

As in the previous case,  $F(m) $ is given by (\ref{F_case_2}).  
The symmetry    $ X ^b _4 $, given in   (\ref{special_3}),  is a divergence symmetry for  
\begin{equation}
\gamma = { 1 \over  2  } 
\end{equation}
with 
\begin{equation}
V^t _4 = 0  , 
\qquad 
V^x _4 = - {  \varepsilon  \over k }  m   . 
\end{equation}
i.e., for this symmetry, there holds 
\begin{equation*}
  X    L + L ( D_t ( \xi^t   ) + D_x ( \xi^x  ) ) =  D_x  \left(  - {  \varepsilon  \over k }  m   \right)   . 
\end{equation*} 
Therefore, we get the  symmetry  $ X ^b _4 $   specified as 
\begin{equation}
{Y} ^b _4 
=  
\left. 
X ^b _4 
\right| _{  \gamma = { 1 \over  2  }  }
=   2 t {\ddt}
+   x {\ddx} 
+ \left(  u  - { x \over k} \right)    {\ddu}
-  { 2 } m   {\ddm} . 
\end{equation}
Using this symmetry,  we obtain the conservation law
\begin{multline}        \label{CL4b} 
D_t \left[ 
 2   t 
\left( 
\varepsilon   m _x   u _x     
+     {  1  \over k}   m  e^{ k  u_x } 
-   { 2 \alpha   \over 3}   m^{3/2}     
\right)
+  x m 
\left(  
u_x   
+  {  1 \over k } 
\right)
- m  u   
\right]  
\\
-  
D_x \left[ 
 2   t 
\left( 
\varepsilon   (   u _t      m _x   +    m_t   u _x  ) 
+  m u_t    e^{ k  u_x }  
  \right) 
+     x 
\left( 
 m u_t  
+   \varepsilon    m _x   u _x     
+   m   u_x    e^{ k  u_x }
+  {  \varepsilon \over k}  m_x    
+   { 2 \alpha   \over 3}   m^{3/2}  
\right)
\right.
\\
\left.
-  u   (   \varepsilon  m_x +   m e^{ k  u_x }  ) 
  +  2  \varepsilon m  u_x 
- {  \varepsilon  \over k }  m   
\right]  
= 0  . 
\end{multline}


\medskip

\noindent {\bf c)} 
  Quadratic   $ H (u_x) =  {  1\over 2}   u_x ^2 $

For the quadratic Hamiltonian,  there are two cases of  $ f(m)  $.

\medskip

\noindent   {\bf 1)}  
Arbitrary  $ f(m)  $

For arbitrary  $ f(m)  $, 
we consider the corresponding function $ F(m) $ as given in (\ref{Arbitrary_f}). 
The symmetry   $ X^c $, given in   (\ref{special_5a}), 
is a divergence symmetry with  
\begin{equation}
V^t = 0   , 
\qquad 
V^x=        - {  \varepsilon  }  m   . 
\end{equation}
This symmetry leads to the conservation law 
\begin{multline}        \label{CL_c} 
D_t \left[ 
 m (  t   u_x   
+  x   ) 
\right]  
\\
- 
D_x \left[ 
  t  
\left( 
  m u_t 
+    \varepsilon     m _x   u _x    
+    {  1  \over 2  }  m u_x ^2 
+  F (m) 
\right)
+ x  (   \varepsilon    m _x   +  m   u_x ) 
-  {  \varepsilon  }  m   
\right]  
= 0 . 
 \end{multline}

\medskip

\noindent {\bf 2)}  Quadratic     $ f(m)  =   \alpha  m^{2} $

Now 
\begin{equation}
F(m)  =   { \alpha  \over 3 }   m^{3} . 
\end{equation}
It is a particular case  of (\ref{F_case_2}). 
The  symmetry 
\begin{equation*}         
{Y} ^c _4  
=  
\left. 
 {Y} ^a _4  
\right| _{p=2}
=   2  t {\ddt}
+  x {\ddx} 
-    m   {\ddm}   , 
\end{equation*}
i.e., symmetry (\ref{special_2_specified}) with $p=2$,  
 is variational. 
It provides a particular case of  conservation law  (\ref{CL4a}) 
corresponding  to  $  \gamma _* = p=  2$, 
namely 
\begin{multline}      \label{CL4a_mod} 
D_t \left[ 
 2  t 
\left( 
\varepsilon     m _x  u _x     
+    {  1\over 2}  m   u_x ^2   
-   { \alpha   \over  3 }   m^{3}     
\right)
+   
 x m u_x  
\right]  
\\ 
- 
D_x \left[ 
  2   t 
\left( 
 \varepsilon   (   u _t      m _x   +    m_t   u _x  ) 
+  m u_t   u_x 
  \right)
+   x 
\left( 
 m u_t 
+    \varepsilon   m _x   u _x    
+   { 1  \over 2 }  m u_x ^2 
+  { \alpha   \over  3 }   m^{3}  
\right)
+  
 \varepsilon   m   u _x     
\right]  
= 0  . 
\end{multline}

The symmetry $ X_ 5 ^c $. given in  (\ref{special_5b}),  is a divergence one with 
\begin{equation}
V^t =  0   , 
\qquad 
V^x=  -   {   \varepsilon   }   x  m        . 
\end{equation}
It gives  the conservation law 
\begin{multline}         \label{CL5a} 
D_t \left[ 
  t ^2 
\left( 
\varepsilon    m _x   u _x    
+    {  1\over 2 }  m   u_x ^2    
-   { \alpha   \over  3  }   m^{3}     
\right)
+ {   t x  } 
  m u_x   
+ 
  { x^2  \over 2 } m  
 -  \varepsilon  t m 
\right]  
\\
- 
D_x \left[ 
      t^2  
\left( 
 \varepsilon   (   u _t      m _x   +    m_t   u _x  ) 
 +   m u_t   u_x 
  \right)  
+  t  x 
\left( 
 m u_t   
+   \varepsilon   m _x  u _x    
+   {  1  \over 2 }  m   u_x ^2  
+ { \alpha   \over 3 }   m^{3}  
\right)
\right.
\\
\left.
  + { x^2  \over 2 }  
(   \varepsilon   m  _x       + m u_x  )  
-    \varepsilon^2  t  m  _x     
- {   \varepsilon   }   x  m    
\right]  
= 0 . 
\end{multline}



The obtained results can be summarized as a theorem.

\begin{theorem}     \label{theorem_conservation} 
Up to equivalence transformations (\ref{transformation_1})--(\ref{transformation_4}),  
system  (\ref{system_mu}),(\ref{Assumptions})
has  the following cases of conservation laws:

For arbitrary $ H (u_x) $ and  arbitrary $ f(m) $,  
the system has three conservation laws 
    (\ref{CLG1}),   (\ref{CLG2}) and  (\ref{CLG3}). 

There are particular cases of $ H(u_x)$ 
which lead to the additional conservation laws. 


\noindent {\bf a)} 
For power $ H (u_x) =  {  1\over p}   u_x ^p  $ 
and   power $ f(m)  =   \alpha  m^{\gamma} $ 
with $  \gamma = \gamma _*= { p \over  2 p -3  }   $, 
there is the additional conservation law    (\ref{CL4a}). 

\noindent {\bf b)} 
For exponential    $ H (u_x)  =  {  1  \over k}    e^{ k  u_x } $ 
and  $ f(m)  =   \alpha  \sqrt{m}   $, 
there  exists the additional conservation law    (\ref{CL4b}). 

\noindent {\bf c)} 
For quadratic   $ H (u_x)  = {  1\over 2}   u_x ^2  $  and arbitrary  $ f(m)  $ 
there is conservation law (\ref{CL_c}).  
If    $ f(m)  =   \alpha  m^{2} $,  then there also hold 
conservation laws    (\ref{CL4a_mod}) and (\ref{CL5a}).
\end{theorem}

Thus, three conservation laws exist for all $ H( u_x) $ and $ f(m)$.
The symmetries which provide additional conservation laws 
 are given  in Table~2.

\begin{equation*}
\begin{array}{|c|c|c|c|}
\hline
   &   
\mbox{arbitrary $f(m)$}   
& 
f(m)  =   \alpha  m^{\gamma} , 
\  
\gamma \neq 2 
& f(m)  =   \alpha  m^{2} 
\\
\hline
\mbox{arbitrary $H  (u_x )$}   &    
 -   &   
-   &        
 - \\
\hline
H  (u_x )  =      {  1\over p}   u_x ^p  , \  p \neq 2    & 
 -   &   
{Y}  ^a _4  &    
- 
\\
\hline
H  (u_x )  =      {  1  \over k}    e^{ k  u_x }  &  
 -  &   
{Y}  ^b _4     &    
- 
\\
\hline
 H  (u_x )  =   {  1  \over 2 } u_x ^2     &  
  X  ^c   &   
X  ^c       &    
X  ^c  ,   {Y} ^c _4 ,  X  ^c _5 
\\
\hline
\end{array}
\end{equation*}
\begin{center}
{\bf Table 2.} Variational and divergence symmetries 
admitted in addition to symmetries  (\ref{basis}), 
which correspond to  arbitrary $H(u_x)$ and  arbitrary $ f(m)$. 
Classification is performed up to equivalence transformations (\ref{transformation_1})--(\ref{transformation_4}).
\end{center}

\section{Interpretation of conservation laws}

\label{Interpretation}

Some conservation laws allow intuitive interpretation.  
Finite group transformations  
\begin{equation*}
t  \rightarrow   \bar{t}    , 
\qquad
 x \rightarrow   \bar{x}      , 
\qquad
u  \rightarrow   \bar{u}    , 
\qquad
m  \rightarrow   \bar{m}   
\end{equation*}
for generators (\ref{generator}) 
solve the ODE system  \cite{bk:Ovsiannikov1978, bk:Ibragimov1985} 
\begin{equation*}
{ d \bar{t} \over d a }   = \xi ^t   (  \bar{t} , \bar{x} , \bar{u} , \bar{m} )    , 
\qquad
{ d \bar{x} \over d a }    = \xi^x     (  \bar{t} , \bar{x} , \bar{u} , \bar{m} )   , 
\qquad
 { d \bar{u} \over d a }   = \eta ^u    (  \bar{t} , \bar{x} , \bar{u} , \bar{m} )  ,  
\qquad
  { d \bar{m} \over d a }   = \eta ^m   (  \bar{t} , \bar{x} , \bar{u} , \bar{m} ) 
\end{equation*}
with the initial conditions 
\begin{equation*}
   \bar{t} | _{a=0} = t    , 
\qquad
   \bar{x}  | _{a=0} = x     , 
\qquad
  \bar{u}   | _{a=0} = u  , 
\qquad
  \bar{m}    | _{a=0} = m  . 
\end{equation*} 
Here  $a$  is the group parameter. 
Its value $a=0$ corresponds to the identity transformation.  
We remark the solutions can be presented  by the exponentiation   \cite{bk:Olver} 
\begin{equation*}
t  \rightarrow   \bar{t}   
= e^{a X} t   , 
\qquad
 x \rightarrow   \bar{x}   
= e^{a X} x   , 
\qquad
u  \rightarrow   \bar{u}   
= e^{a X} u   , 
\qquad
m  \rightarrow   \bar{m}   
= e^{a X} m   .   
\end{equation*}

Specifically,  we  consider a particular case of the mean field games 
described in Appendix   \ref{HJB-KE}. 
Let the dynamics of the agents be described by the SDE 
\begin{equation}
d X_s =   \alpha  _s   d s   + \sqrt{2 \varepsilon} d B_s  , 
\end{equation}
where $  \alpha  _s  $ is the agent's control. 
Brownian motions $ B_s $  for different agents are assumed to be independent. 
Each  agent minimizes 
\begin{equation}
\mbox{E} 
\left[
\int _ 0 ^T 
\left(  {  \alpha  _s  ^2 \over 2 } + f(m (s, X_s ) ) \right)  ds   
+ G ( X_T , m(T, X_T ) )  
\ 
| 
\ 
X_0 = x
\right]  . 
\end{equation}  
The running costs have two terms. 
The first term  $ {  \alpha  _s  ^2 \over 2 } $ requires to minimise the state changes. 
The second term $ f(m) $  represents  the interaction of the agents. 
It can be a willingness to avoid others if $ f(m) $ is an increasing function
or a willingness to be like others if $ f(m) $ is a decreasing function.

We obtain the Hamiltonian 
\begin{equation*}
H ( x, p ,  m )  = { p ^2   \over 2 }   -  f (m)   , 
\end{equation*}
the MFG system 
\begin{subequations}     \label{example_system}
\begin{gather}   
 - u_t  -  \varepsilon   u _{xx}  +   { u_x ^2    \over 2 }     =   f ( m )    , 
\\
 m_t  -  \varepsilon  m _{xx}   -  ( m    u_x ) _x         = 0  . 
\end{gather}
\end{subequations}
and the optimal feedback control 
\begin{equation}    \label{control}
{  \alpha  _s } = -  u_x  (t,x)   . 
\end{equation}

System (\ref{example_system}) has four conservation laws. 
Conservation law   (\ref{CLG2}) corresponds to the symmetry $ X_2 $, 
which represents the translation of the state variable $x$
\begin{equation*}
   \bar{t}   
=   t   , 
\qquad
  \bar{x}   
=  x   + a , 
\qquad
  \bar{u}   
= u   , 
\qquad
  \bar{m}   
= m   .
\end{equation*}
Taking (\ref{control}) into account, 
it is possible to consider   this  conservation law as the conservation of the average control value.  
In mechanics, translations of the spatial coordinates lead to  the conservation of momenta.

Conservation law (\ref{CLG3}) is provided by translations of $u$, 
\begin{equation*}
   \bar{t}   
=   t   , 
\qquad
  \bar{x}   
=  x   , 
\qquad
  \bar{u}   
= u  + a  , 
\qquad
  \bar{m}   
= m  , 
\end{equation*}
given by the symmetry $X_3$.  
It  has the same interpretation for the quadratic Hamiltonian as for the arbitrary  Hamiltonian. 
It represents the conservation of the probability measure  (\ref{normalization}).  
It is analogous to the conservation of mass in mechanical systems.

Conservation law   (\ref{CL_c}) is derived with the help of  the Galileo-type symmetry $X^c$. 
This symmetry generates the group transformation   
\begin{equation*}
   \bar{t}   
=   t   , 
\qquad
  \bar{x}   
=  x   + a t  , 
\qquad
  \bar{u}   
= u -  a x - { a^2 \over 2 } t    , 
\qquad
  \bar{m}   
= m   .
\end{equation*}
It  can be viewed as the law describing the change of the average state value.  
In mechanics, the  Galileo transformation corresponds to the law of motion of the centre of mass.

There seems to be no natural interpretation for conservation law (\ref{CLG1}) 
as well as for the additional  conservation laws  (\ref{CL4a_mod}) and  (\ref{CL5a}), 
which exist for quadratic $ f(m) $.    
These conservation laws correspond to the time translation
\begin{equation*}
   \bar{t}   
=   t  + a  , 
\qquad
  \bar{x}   
=  x   , 
\qquad
  \bar{u}   
= u   , 
\qquad
  \bar{m}   
= m   , 
\end{equation*}
the scaling 
\begin{equation*}
   \bar{t}   
= e^{2a}   t   , 
\qquad
  \bar{x}   
=   e^{a}  x    , 
\qquad
  \bar{u}   
= u   , 
\qquad
  \bar{m}   
=  e^{-a} m   , 
\end{equation*}
and the projective transformation in  $ (t , x ) $ plane 
\begin{equation*}
   \bar{t}   
=   { t  \over 1 - a t }   , 
\qquad
  \bar{x}   
=  { x  \over 1 - a t }   
\end{equation*}
supplemented with the  transformation of the dependent variables 
\begin{equation*}
   \bar{u}   
= u 
-     { a x ^2  \over 2(  1 - a t )  }  
-  \varepsilon \ln (  1 - a t )     , 
\qquad
  \bar{m}   
= (  1 - a t )  m   .
\end{equation*}
The transformations are generated by symmetries
$ X_1 $,  $ Y ^c _4 $  and $ X ^c _5 $,  respectively. 
In mechanics, the time translation provides  the conservation of energy.  
Conservation laws corresponding to the scaling transformations 
and the projective transformations 
are known in   mechanics,  kinetic theory  gases,  hydrodynamics, etc. \cite{Bobylev1989}.  
They have no specific names.

\section{Conclusion}

\label{Conclusion}

The paper considers  system    (\ref{system_mu}),(\ref{Assumptions}). 
This system is the simplest case of the mean field games system   (\ref{separable_system}), 
which has a variational formulation due to the separable Hamiltonian. 
System    (\ref{system_mu})  corresponds to state-independent  Hamiltonians 
in one-dimensional state space. 
The paper provides Lie point symmetry analysis of the system. 
Then,  the variational and divergence symmetries are selected. 
They are used for the  application of the Noether theorem to derive the conservation laws. 
The presented  approach  provides the symmetries  and  conservation laws 
for the general case of  the system    (\ref{system_mu}),(\ref{Assumptions}) 
with arbitrary functions $ H( u_x)$  and $ f(m)$ 
and identifies the particular cases of $ H( u_x)$  and $ f(m)$ 
which lead to the additional symmetries and conservation laws.  
Some conservation laws have natural interpretations.

Conservation laws of the first-order mean field games systems, 
which correspond to   parameter  $ \varepsilon = 0$ in (\ref{general_system}), 
were considered in \cite{Gomes2016, Gomes2018}.  
To our  knowledge, there have been no attempts to investigate 
second-order  mean field games equations for conservation laws 
that makes the results of this paper original. 
The study given here  can also be  performed 
for more general MFG systems, such as

\begin{itemize}

\item 

general system   (\ref{separable_system}) 
for  multi-dimensional state space with  state-dependent  $H$ and $f$, 

\item 

system  (\ref{x_independent}) 
for multi-dimensional state space with  state-independent  $H$ and $f$, 

\item 

system   (\ref{one_dim_system}) 
for one-dimensional state space with  state-dependent  $H$ and $f$. 

\end{itemize}
It is also possible to consider stationary  versions of these mean field games equations.


For example, system   (\ref{separable_system}) with   arbitrary  $H(x,  \nabla u)$ and $f(x, m)$ 
admits only two symmetries 
\begin{equation*}  
X_1 = {\ddt}  , 
\qquad 
X_2 = {\ddu}      . 
\end{equation*}
The generator $  X_1  $ is a variational symmetry  
and the generator  $  X_2  $  is a divergence symmetry 
of the Lagrangian function  (\ref{separable_Lagrangian}).  
They provide the conservation laws 
\begin{equation*}    
D_t \left[ 
   \varepsilon     \nabla      m     \cdot      \nabla    u   
+ m  H (  x,  \nabla  u  ) 
- F  ( x, m )  
\right]
- 
\mbox{div}  \left[ 
   \varepsilon   (   u _t     \nabla   m   + m_t    \nabla    u      ) 
+  m u_t  H _{\nabla  u }   (  x,  \nabla  u    ) 
\right] 
= 0   
\end{equation*}
and 
\begin{equation*}      
- D_t \left[ 
   m  
\right]  
+ 
\mbox{div}
\left[ 
\varepsilon   \nabla m     
+  m  H _  { \nabla  u }  ( x ,  \nabla u  ) 
\right]  
= 0 , 
\end{equation*}
respectively.  
These conservation laws generalize  
conservation laws  (\ref{CLG1}) and  (\ref{CLG3}), 
which hold for system   (\ref{system_mu})   
with arbitrary  $H(u_x)$ and $f(m)$.

Conservation laws reflect fundamental properties of the differential equations, 
that  helps to understand and analyse the models.  
For this reason, it is crucial to preserve them in numerical simulation \cite{Hairer}. 
In particular, 
conservation of measure   (\ref{normalization}) 
is always respected by numerical methods  for the MFG equations \cite{Achdou2020}.  
Finally,  it is worth mentioning that 
symmetries of PDEs can be used to find particular solutions 
\cite{bk:Ovsiannikov1978, bk:Ibragimov1985, bk:Olver}.   
This  has  not been explored  for the mean field games equations.

\section*{Acknowledgments}

The author is grateful to   Prof.   Vladimir  A.  Dorodnitsyn 
for discussions concerning  methods of Lie group analysis.

\section*{Appendices}

\appendix

\section{A sketch of the derivation of the MFG system}

\label{HJB-KE}


In this Appendix, we sketch the derivation of the system    (\ref{general_system}), 
which describes Nash equilibrium in a game  with a continuum of agents. 
The presentation follows \cite{Cardaliaguet2020}. 
We refer to this book chapter for details 
and to  \cite{Fleming2012, Yong1999} for the optimal control theory.

Each agent exercises control $ \alpha $ 
in the stochastic differential equation  in $  \mathbf{R}^d $
\begin{equation}   \label{sde}
d X _s = b ( X _s , \alpha _s ,  m(s, X_s ) ) ds  
+ \sqrt{2 \varepsilon} d B_s , 
\end{equation}    
where $ B_s $ is a Brownian motion.    
Here $ m( t, x ) $ is the density of the probability measure in $  \mathbf{R}^d $, 
which describes the distribution of the agents.   
The agent minimizes 
\begin{equation}
\mbox{E} 
\left[
\int _ 0 ^T 
L (  X _s , \alpha _s ,  m(s, X_s ) ) ds   
+ G ( X_T , m(T, X_T ) )  
\ 
| 
\ 
X_0 = x
\right] , 
\end{equation}  
where  $L (x, \alpha, m) $ and $G (x,m)$ are running and terminal costs. 
They depend on the agent's state $x$ and the distribution of agents $m$. 
The running cost $L$ also depends on the agent's control $ \alpha$.

This problem provides the value function 
\begin{equation}
u ( t,x ) 
= 
 \mathop{\mbox{inf}} \limits_{\alpha}
\mbox{E} 
\left[
\int _ t ^T 
L (  X _s , \alpha _s ,  m(s, X_s ) ) ds   
+ G ( X_T , m(T, X_T) )
\ 
| 
\ 
X_t = x
\right] 
\end{equation}  
The infimum is taken over admissible controls.   
Function  $X _s$ satisfies SDE   (\ref{sde}) 
with the initial condition $ X_t = x$.

Each agent makes an infinitesimally small contribution 
to the population dynamics. 
Therefore, the agent gets a problem of individual optimization 
with given time dependent  distribution of the other agents. 
For the value function, we get PDE    (\ref{general_equation_1}) 
with 
\begin{equation}
H ( x, p, m ) 
= 
 \mathop{\mbox{sup}} \limits_{\alpha} 
[ - b ( x, \alpha , m ) \cdot  p - L  ( x, \alpha , m ) ]  . 
\end{equation}  
The optimal control has the feedback form 
$ \alpha ^* =  \alpha ^* ( t,x) $ 
and 
\begin{equation}
b ( x,\alpha ^* ( t,x)  , m (t) ) 
= - H_p  ( x,  \nabla   u  , m (t) )   . 
\end{equation}

If the agents exercise the optimal behaviour 
and their noises $ B_s$  are independent, 
the evolution of the agents' distribution is given 
by the forward Kolmogorov equations (\ref{general_equation_2})


\section{Preliminary analysis of the determining equations}

\label{preliminary}

Preliminary analysis  identifies cases of the Hamiltonian function  $ H(u_x)$, 
which should  be considered.

Lie point symmetries 
\begin{equation}    \label{A_generator}
X 
=  \xi ^t  {\ddt} 
+  \xi ^x  {\ddx} 
+  \eta  ^u  {\ddu}  
+  \eta  ^m  {\ddm}  
\end{equation}
of the MFG systen 
\begin{subequations}     \label{A_system}
\begin{gather}     \label{A_equation_u}
F_1 = - u_t  -  \varepsilon   u _{xx}  + H ( u_x ) = f ( m )  , 
\\
  \label{A_equation_m}
F_2  =     m_t  -  \varepsilon  m _{xx}   -   m _x H '  ( u_x )     -      m H ''  ( u_x )  u _{xx}     = 0 
\end{gather}
\end{subequations}
are specified by the determining  equations 
\begin{subequations}         \label{determine} 
\begin{gather}
E_1  = 
\left. 
X ( F_1  )
\right| _{  F_1 = 0 , F_2 =0 } 
= 0 , 
\\
E_2 = 
 \left. 
 X (  F_2   )
 \right| _{  F_1 = 0 , F_2 =0 } 
= 0  .
\end{gather}
\end{subequations}  
As discussed in Section  \ref{Symmetries}, 
the generator $X$ is assumed to be prolonged to all variables 
involved in the equations.

Application of the prolonged generator $X$ to the equations  
$ F_1= 0$ and $ F_2= 0$ gives the system of two PDEs 
for the coefficients of the generator $X$. 
The generator coefficients depend on the independent   and dependent variables $ t $, $x$, $u$ and $ m$. 
The system also  depends on the functions $ H ( u_x) $ and $f(m)$ and their derivatives. 
Overall, the system depends on the variables  
$ \{  t, x, u , m , u_t , m_t,   u_x , m_x,   u_{tx} , m_{tx} , u_{xx} , m_{xx}  \} $.  
Since the action of the prolonged generator $X$ to the MFG equations  
 is considered on the solutions of the MFG equations, we eliminate two of these variables.   
Substituting 
\begin{equation}    \label{substitution_m}
  m _{xx} 
= 
{ 1 \over \varepsilon  } 
(     m_t     -   m _x H '  ( u_x )     -      m H ''  ( u_x )  u _{xx}   ) 
\end{equation}
and then 
\begin{equation}    \label{substitution_u}
   u _{xx} 
=   
{ 1 \over \varepsilon  } 
(  - u_t   + H ( u_x ) -  f ( m )  )     , 
\end{equation}
we eliminate  $  u_{xx} $  and   $  m_{xx}  $.  
These  substitutions lead to the equations  
\begin{subequations}
\begin{gather}
E_1  ( t,x, u , m , u_t , m_t,   u_x , m_x,   u_{tx} , m_{tx}) 
= 
\left. 
X ( F_1  )
\right| _{   (\ref{substitution_m}) ,  (\ref{substitution_u})  } 
= 0 , 
\\
E_2  ( t,x, u , m , u_t , m_t,   u_x , m_x,   u_{tx} , m_{tx}) 
= 
\left. 
X ( F_2   )
\right| _{   (\ref{substitution_m}) ,  (\ref{substitution_u})     } = 0   . 
\end{gather}
\end{subequations}
In detail, these equations take the form 
\begin{equation*}      
E_1 = 
\left. 
\left(
 -  \zeta _t  ^u  
-  \varepsilon    \zeta _{xx}  ^u  + 
H' (  u_x )  \zeta _x  ^u 
 -  f '  ( m )  
+  \eta  ^m  
\right) 
\right| _{   (\ref{substitution_m}) ,  (\ref{substitution_u})  } 
= 0 , 
\end{equation*}      
\begin{multline*}      
E_2 = 
\left(
\zeta _t  ^m
 -  \varepsilon  \zeta _{xx} ^m  
-  \zeta _x  ^m    H '  ( u_x )  
-   m _x   H ''  ( u_x ) \zeta _x  ^u    
 \right.
 \\ 
 \left. 
\left. 
-    \eta  ^m  H ''  ( u_x )  u _{xx}
- m  H '''  ( u_x )  \zeta _x  ^u    u _{xx}
-   m H ''  ( u_x )  \zeta _{xx}  ^u 
\right) 
\right| _{   (\ref{substitution_m}) ,  (\ref{substitution_u})  } 
= 0  .
\end{multline*}  
Clearly, these equations can be split for  
$ u_t $, $m_t$,  $m_x$,   $u_{tx}$ and  $ m_{tx}$ 
because the generator coefficients,    $ H(u_x)$ and  $ f(m)$ 
do not depend on these variables. 


We proceed as follows.

\begin{enumerate}

\item 

From 
\begin{equation*}   
{ \partial E_1 \over \partial u_{tx} } 
= 2  \varepsilon 
(  \xi ^t   _x +     \xi ^t _ u  u_x  +   \xi ^t _ m   m_x  ) = 0 , 
\end{equation*}
where $  \xi ^t  =  \xi ^t  (  t,x,u,m) $, 
we obtain 
\begin{equation*}   
 \xi ^t  =  A   (t)  , 
\end{equation*} 
where $ A(t) $ is an arbitrary function. 

\item 

The equation
\begin{equation*}   
{ \partial ^2 E_1 \over \partial m_{x} ^2  } 
= 2  \varepsilon 
(  \xi ^x   _{mm}  u_x   -     \eta ^u   _{mm}    )    = 0 
\end{equation*}
for    $  \xi ^x  =  \xi ^x  (  t,x,u,m) $ and  $  \eta  ^u  =  \eta  ^ u (  t,x,u,m) $ 
gives 
\begin{equation*}   
 \xi ^x =   B   _1   (  t,x,u )  m    +    B   _0   (  t,x,u )    
\end{equation*} 
and
\begin{equation*}   
 \eta  ^u  =   C   _1   (  t,x,u )  m    +    C   _0   (  t,x,u )    . 
\end{equation*}

\item 

The equation
\begin{equation*}   
{ \partial ^4 E_2 \over \partial u_{x} ^2    \partial m_{x} ^2   } 
= 2  ( B   _1   (  t,x,u )   u_x   -   C   _1   (  t,x,u ) ) H '''' (u_x ) = 0
\end{equation*}
identifies the special case of the Hamiltonian function 
\begin{equation*}    
 H  (u_x )  = h_3    u_x ^3  +  h_2  u_x ^2  +  h_1  u_x   +  h_0 . 
\end{equation*}
This case will  be treated separately. 
It is  convenient  to split this case into two: 
the case of the cubic Hamiltonian 
\begin{equation}    \label{A_form_1a}
 H  (u_x )  = h   u_x ^3  +  h_2  u_x ^2  +  h_1  u_x   +  h_0 , 
\qquad 
h \neq 0
\end{equation}
and the case of the quadratic Hamiltonian
\begin{equation}    \label{A_form_1b}
 H  (u_x )  = h    u_x ^2  +  h_1  u_x   +  h_0 , 
\qquad 
h \neq 0  . 
\end{equation}
Linear Hamiltonians are excluded by requirement  (\ref{Assumptions}).

We  continue with the assumption that the Hamiltonian has a form 
different from   (\ref{A_form_1a}) and (\ref{A_form_1b}). 
It follows  
\begin{equation*}   
 B   _1   (  t,x,u )  = 0 , 
\qquad  
C   _1   (  t,x,u )  = 0 . 
\end{equation*}

\item 

The equation
\begin{equation*}   
{ \partial ^2 E_2 \over \partial m_{x} ^2      } 
= - 2     \varepsilon  \eta ^m   _{mm}       = 0 
\end{equation*}
for   $  \eta  ^m  =  \eta  ^m (  t,x,u,m) $  
provides 
\begin{equation*}   
 \eta  ^m  =   G   _1   (  t,x,u )  m    +    G  _0   (  t,x,u )     . 
\end{equation*}

\item 

From 
\begin{equation*}   
{ \partial ^2 E_2 \over \partial u_{x}  \partial m_{t}     } 
=  2   ( B   _0   (  t,x,u )  ) _u         = 0 
\end{equation*}
we get 
\begin{equation*}   
B   _0   (  t,x,u ) =   \tilde{B}   _0   (  t,x ) . 
\end{equation*}

\end{enumerate}

After all these simplifying steps,  
we obtain the symmetry coefficients in the form 
\begin{equation*}   
\xi ^t = A(t)  ,   
\qquad 
\xi ^x =\tilde{B}   _0   (  t,x ) , 
\qquad 
\eta ^u  = C_0 (t, x, u) , 
\qquad 
\eta  ^m  = G_1 (t, x, u) m + G_0 (t, x, u) . 
\end{equation*}
These such   coefficients 
we obtain  the equation 
\begin{equation*}   
{ \partial ^3 E_2 \over \partial u_{x} ^2     \partial m_{x}    } 
=   ( F_1 ( t,x,u)  u_x +  F_2 ( t,x,u)  )   H '''' (u_x ) +  F_3 ( t,x,u)  H ''' (u_x ) = 0   , 
\end{equation*}
where functions $ F_1 $, $ F_2 $  and $ F_3 $ depend on the derivatives of the functions 
 $\tilde{B}   _0   (  t,x ) $ and   $C_0 (t, x, u)$.   
This equation is {\it  the classifying equation}  for the Hamiltonian function. 
In addition to the cases (\ref{A_form_1a})  and  (\ref{A_form_1b})     found earlier, 
we get two more cases
\begin{equation}     \label{A_form_2}
 H  (u_x )  =      h  (  u_x + q ) ^p     +  h_2  u_x ^2  +  h_1  u_x   +  h_0  , 
\qquad 
h \neq 0 ,  
\qquad 
p  \neq 0, 1, 2, 3 
\end{equation}
and 
\begin{equation}     \label{A_form_3}
 H  (u_x )  =      h   e^{ k  u_x }   +  h_2  u_x ^2  +  h_1  u_x   +  h_0  , 
\qquad 
h \neq 0 ,  
\qquad 
k  \neq 0 . 
\end{equation}
Further examination of the determining equations 
does not provide any other special cases of the Hamiltonian function. 
Thus, in addition to the special cases 
(\ref{A_form_1a}),    (\ref{A_form_1b}),    (\ref{A_form_2})  and  (\ref{A_form_3}), 
we have only the general case of  the arbitrary $ H (u_x) $.

This is the end of the preliminary analysis of the determining equations.  
After this we proceed with the specified forms  of the Hamiltonian function. 
We should return to the determining  equations (\ref{determine}) 
 and solve them for the different cases of $H (u_x) $. 
It provides the admitted symmetries and the special cases of function $f(m)$.


\begin{thebibliography}{00}

\bibitem{Lasry_Lions_2006a}
J.-M. Lasry and  P.-L. Lions, 
Mean field games. I - The stationary case | 
Jeux à champ moyen. I - Le cas stationnaire, 
{\it  Comptes Rendus Mathematique}, 
{\bf 343}(9),  619--625, 
2006

\bibitem{Lasry_Lions_2006b}
J.-M. Lasry and  P.-L. Lions,  
Mean field games. II - Finite horizon and optimal control 
| Jeux à champ moyen. II - Horizon fini et contrôle optimal, 
{\it  Comptes Rendus Mathematique}, 
{\bf 343}(10),  679--684, 
2006

\bibitem{Lasry_Lions_2007}
J.-M. Lasry and  P.-L. Lions,  
Mean field games, 
{\it Japanese Journal of Mathematics}, 
{\bf 2}(1),  229--260, 
2007

\bibitem{Huang_2006}
M. Huang, R. P. Malhamé and P. E. Caines, 
Large population stochastic dynamic games: closed-loop Mckean-Vlasov systems 
and the Nash certainty equivalence principle. 
{\it Commun. Inf. Syst.} 
{\bf 6}, 221--252, 
2006


\bibitem{Huang_2007b} 
M. Huang, R. P. Malhamé and P. E. Caines, 
Large-population cost-coupled LQG problems with nonuniform agents: 
individual-mass behavior and decentralized $\epsilon$-Nash equilibria, 
{\it  IEEE Trans. Autom. Control}, 
{\bf 52}, 1560--1571, 
2007


\bibitem{Aumann_1964} 
R. J. Aumann, 
Markets with a continuum of traders. 
{\it Econometrica},  
{\bf 32} (1-2), 39--50, 
1964







\bibitem{GLL2011}
 O. Guéant, J.-M. Lasry and  P.-L. Lions,   
Mean field games and applications, 
Lecture Notes in Mathematics 2003, 
205--266, 2011. 


\bibitem{Sircar2015} 
P. Chan and R.  Sircar,  
Bertrand and Cournot Mean Field Games, 
{\it   Applied Mathematics and Optimization},  
{\bf 71}(3),  533--569, 
2015

\bibitem{Graber2018}
P. J. Graber and  C. Mouzouni, 
Variational Mean Field Games for Market Competition, 
{\it 	Springer INdAM Series}, 
{\bf 28},  93--114, 
2018 

\bibitem{Carmona2021} 
R. Carmona and G. Dayanlkll, 
	Mean Field Game Model for an Advertising Competition in a Duopoly, 
 	{\it International Game Theory Review}, 
{\bf 23}(4), 2150024, 
	2021

\bibitem{Chenavaz2021} 
R. Chenavaz, C., Paraschiv and G. Turinici, 
	Dynamic Pricing of New Products in Competitive Markets: A Mean-Field Game Approach, 
{\it  	Dynamic Games and Applications}, 
{\bf 11}(3), 463--490, 
2021

\bibitem{Gomes2022} 
D. A. Gomes  and   J.  Saúde, 
	A Mean-Field Game Approach to Price Formation, 
{\it 	Dynamic Games and Applications}, 
{\bf 11}(1),  29--53, 
2021 	

\bibitem{Gomes2023}
D. Gomes,   J.  Gutierrez and   R.  Ribeiro,  
A Random Supply Mean Field Game Price Model, 
{\it	SIAM Journal on Financial Mathematics}, 
{\bf 14}(1),  188--222, 
2023 

\bibitem{Carmona2015} 
R. Carmona,  J.-P.  Fouque  and L.-H. Sun, 
	Mean field games and systemic risk, 
{\it 	Communications in Mathematical Sciences}, 
{\bf 13}(4),  911--933, 
2015 

\bibitem{Carmona2017} 
R. Carmona, F. Delarue and D. Lacker, 
Mean Field Games of Timing and Models for Bank Runs, 
{\it Applied Mathematics and Optimization},  
{\bf 76}(1), 217--260, 
2017

\bibitem{Fu2021} 
G. Fu,   P.  Graewe,   U.   Horst  and A.   Popier, 
	A mean field game of optimal portfolio liquidation, 
{\it  	Mathematics of Operations Research}, 
{\bf 46}(4),  1250--1281, 
2021

\bibitem{Fu2023a} 
G. Fu and   C. Zhou, 
	Mean field portfolio games, 
{\it 	Finance and Stochastics}, 
{\bf 27}(1),  189--231, 
2023

\bibitem{Fu2023b} 
G. Fu, 
	Mean field portfolio games with consumption, 
{\it 	Mathematics and Financial Economics}, 
{\bf 17}(1), 79--99, 
2023 

\bibitem{Lachapelle2016}  
A. Lachapelle,    J.-M.  Lasry,     C.-A.    Lehalle  and  P.-L.   Lions,  
Efficiency of the price formation process in presence of high frequency participants: 
a mean field game analysis
{\it 	Mathematics and Financial Economics}, 
{\bf 10}(3), 223--262, 
2016 

\bibitem{Wolfram2011} 
A. Lachapelle  and M.-T. Wolfram, 
	On a mean field game approach modelling congestion and aversion in pedestrian crowds, 
{\it 	Transportation Research Part B: Methodological}, 
{\bf 45}(10), 1572--1589, 
2011

\bibitem{Wolfram2013} 
M. Burger, M.  Di Francesco, P. A.   Markowich and  M.-T.  Wolfram, 
	On a mean field game optimal control approach modelling fast exit scenarios in human crowds, 
{\it 	Proceedings of the IEEE Conference on Decision and Control}, 
6760360,  3128--3133, 
2013 

\bibitem{Wolfram2014} 
M. Burger, M.  Di Francesco, P. A.   Markowich and  M.-T.  Wolfram, 
	Mean field games with {non}linear mobilities in pedestrian dynamics, 
{\it  	Discrete and Continuous Dynamical Systems - Series B}, 
{\bf 19}(5),  1311--1333, 
2014 

\bibitem{Achdou2019} 
Y.  	Achdou  and J.-M. 	Lasry,  
Mean field games for modelling crowd motion, 
{\it    Computational Methods in Applied Sciences}, 
{\bf 47},  17--42, 
2019

\bibitem{Wolfram2016} 
M. Burger, A. Lorz and M.-T. Wolfram, 
On a Boltzmann mean field model for knowledge growth, 
{\it SIAM Journal on Applied Mathematics}, 
{\bf  76}(5),  1799--1818, 
2016

\bibitem{Sircar2017} 
P. Chan and R.  Sircar, 
	Fracking, renewables, and mean field games, 
 	{\it 	SIAM Review}, 
{\bf 59}(3),  588--615, 
2017

\bibitem{Alasseur2020} 
C. Alasseur,   I.  Ben Taher  and A.   Matoussi,  
An Extended Mean Field Game for Storage in Smart Grids, 
{\it 	Journal of Optimization Theory and Applications}, 
{\bf 184}(2),  644--670, 
2020 

\bibitem{Djehiche2020} 
B. Djehiche, J.  Barreiro-Gomez  and H. Tembine, 
	Price Dynamics for Electricity in Smart Grid Via Mean-Field-Type Games, 
{\it 	Dynamic Games and Applications}, 
{\bf 10}(4),   798--818, 
2020 

\bibitem{Tankov2021} 
R. Aïd,   R.  Dumitrescu  and P.  Tankov, 
The entry and exit game in the electricity markets: a mean-field game approach, 
{\it  	Journal of Dynamics and Games}, 
{\bf 8}(4),   331--358, 
2021 

\bibitem{Carmona2022} 
R. Carmona,   G.  Dayanıklı   and M.  Laurière, 
Mean Field Models to Regulate Carbon Emissions in Electricity Production, 
{\it Dynamic Games and Applications}, 
{\bf 12}(3), 897--928, 
2022

\bibitem{Achdout2016} 
Y. Achdou, P.-N. Giraud,  J.-M.  Lasry   and P.-L.  Lions, 
	A Long-Term Mathematical Model for Mining Industries, 
{\it 	Applied Mathematics and Optimization}, 
{\bf 74}(3), 579--618, 
2016 

\bibitem{Achdout2022a} 
Y. Achdou,  C.  Bertucci,   J.-M.  Lasry,   P.-L. Lions,   A.  Rostand  and J. A.  Scheinkman, 
	A class of short-term models for the oil industry that accounts for speculative oil storage, 
{\it Finance and Stochastics}, 
{\bf 26}(3),  631--669, 
2022 

\bibitem{Gomes2014} 
D.A. Gomes  and J. Saúde, 
Mean Field Games Models - A Brief Survey,  
{\it 	Dynamic Games and Applications}, 
{\bf 4}(2),  110--154, 
2014 

\bibitem{Huang_2017}
M. Huang, R. P. Malhamé and P. E. Caines, 
Mean Field Games, 
In  T.   Başar and  G.  Zaccour  (eds) 
Handbook of Dynamic Game Theory, 
Springer,
2018

\bibitem{Carmona2018a}
R. Carmona and F. Delarue, 
{\it 	Probabilistic Theory of Mean Field Games with Applications I: 
Mean Field FBSDEs, Control, and Games}, 
Probability Theory and Stochastic Modelling 83,  
 Springer, 
2018 	

\bibitem{Carmona2018b} 
R. Carmona and F. Delarue, 
{\it Probabilistic Theory of Mean Field Games with Applications II: 
Mean Field Games with Common Noise and Master Equations}
Probability Theory and Stochastic Modelling 84, 
Springer, 
2018

\bibitem{Cardaliaguet2019}
P. Cardaliaguet,  F.  Delarue,  J.-M. Lasry and  P.-L. Lions,    
{\it The master equation and the convergence problem in mean field games},   
Annals of Mathematics Studies 201, 
Princeton University Press, 
2019

\bibitem{Cardaliaguet2020}
P. Cardaliaguet and A. Porretta,  
	An introduction to mean field game theory, 
in  	P. Cardaliaguet and A. Porretta (Eds.) 
Mean Field Games, 
	Lecture Notes in Mathematics {\bf 2281} 
(Cetraro, Italy 2019), 
pp. 1-- 158, 
2020










\bibitem{Naz2008} 
R. Naz, F. M. Mahomed and D. P. Mason, 
Comparison of different approaches to conservation laws 
for some partial differential equations in fluid mechanics, 
{\it Appl. Math. Comput.}, 
{\bf 205} (1),   212--230, 
2008

\bibitem{bk:Noether1918}
E.~Noether, 
Invariante variations problem.
{\it Konigliche Gesellschaft der Wissenschaften zu Gottingen,  
Nachrichten, Mathematisch-Physikalische Klasse Heft 2}, pages 235--257, 1918.
English translation: Transport Theory and Statist. Phys., 1(3), 1971,   183--207

\bibitem{bk:Ovsiannikov1978}
L.~V. Ovsiannikov, 
{\it Group Analysis of Differential Equations}, 
Nauka, Moscow, 1978.
English translation, {W}.~{F}. {A}mes,  Ed., 
published by Academic   Press, New York, 
1982

\bibitem{bk:Ibragimov1985}
N.~H. Ibragimov, 
{\it Transformation Groups Applied to Mathematical Physics}, 
Reidel, Boston, 
1985

\bibitem{bk:Olver}
P.~J. Olver, 
{\it Applications of Lie Groups to Differential Equations}, 
Springer, New York, 
1986

\bibitem{Bessel-Hagen}
E.   Bessel-Hagen,  
Über die Erhaltungssätze der Elektrodynamik, 
{\it Math. Ann.}, 
{\bf 84}(3-4), 258--276, 
1921

\bibitem{Bobylev1989}
A. V. Bobylev and N. Kh. Ibragimov, 
Interconnectivity of symmetry properties for equations of dynamics, kinetic theory of gases, and hydrodynamics, 
{\it Matem. Mod.}, 
{\bf 1}(3), 100--109, 
1989

\bibitem{Gomes2016}
D. A.  Gomes,   L. Nurbekyan and    M. 	 Sedjro, 
One-dimensional forward–forward mean-field games, 
{\it Applied Mathematics and Optimization}, 
{\bf 74}(3),  619--642, 
2016 

\bibitem{Gomes2018}
D. Gomes, L.  Nurbekyan  and  M.  Sedjro, 
	Conservation laws arising in the study of forward–forward mean-field games,  
{\it 	Springer Proceedings in Mathematics and Statistics}, 
{\bf 236},   643--649, 
2018 

\bibitem{Hairer}  
E. Hairer, C. Lubich and G. Wanner, 
{\it Geometric Numerical Integration: 
Structure-Preserving Algorithms for Ordinary Differential Equations}, 
Springer, 
2006

\bibitem{Achdou2020}
Y. Achdou and   M.  Laurière, 
Mean field games and applications: Numerical aspects, 
in  	P. Cardaliaguet and A. Porretta (Eds.) 
Mean Field Games, 
	Lecture Notes in Mathematics {\bf 2281} 
(Cetraro, Italy 2019), 
pp. 249--307, 
2020

\bibitem{Fleming2012}
 W. H. Fleming and R. W. Rishel, 
{\it Deterministic and Stochastic Optimal Control}, 
Springer Science \& Business Media, Berlin, 2012

\bibitem{Yong1999}
J. Yong and X. Y. Zhou, 
{\it Stochastic controls: Hamiltonian systems and HJB equations}, 
Springer Science \& Business Media, Berlin, 
1999





\end{thebibliography}
\end{document}